\documentclass[pra,superscriptaddress,aps]{revtex4}
\usepackage{amsmath,epsfig,amsfonts,graphicx,hyperref}
\usepackage{subfigure}
\usepackage{graphicx}
\usepackage[american]{babel}
\usepackage{amssymb}
\usepackage{amsfonts}
\usepackage{color}
\usepackage[normalem]{ulem}

\begin{document}
\newcommand{\be}{\begin{equation}}
\newcommand{\ee}{\end{equation}}
\newtheorem{corollary}{Corollary}[section]
\newtheorem{remark}{Remark}[section]
\newtheorem{definition}{Definition}[section]
\newtheorem{theorem}{Theorem}[section]
\newtheorem{proposition}{Proposition}[section]
\newtheorem{lemma}{Lemma}[section]
\newtheorem{help1}{Example}[section]
\title{
Spatiotemporal algebraically localized waveforms
for a nonlinear Schr{\"o}dinger model with gain and loss
}

\author{Z. A. Anastassi}
\affiliation{Department of Mathematics, Statistics and Physics, College of Arts and
Sciences, Qatar University, P. O. Box 2713, Doha, Qatar}
\author{G. Fotopoulos}
\affiliation{Department of Mathematics, Statistics and Physics, College of Arts and
Sciences, Qatar University, P. O. Box 2713, Doha, Qatar}
\author{D. J. Frantzeskakis}
\affiliation{Department of Physics,
National and Kapodistrian University of Athens, Panepistimiopolis, Zografos, Athens
15784, Greece}
\author{T. P. Horikis}
\affiliation{Department of Mathematics, University of Ioannina, Ioannina 45110,
Greece}
\author{N. I. Karachalios}
\affiliation{Department of Mathematics, University of the Aegean, Karlovassi, 83200
Samos, Greece}
\author{P. G. Kevrekidis}
\affiliation{Department of Mathematics and Statistics,
University of Massachusetts, Amherst MA 01003-4515, USA}
\author{I. G. Stratis}
\affiliation{Department of Mathematics,
National and Kapodistrian University of Athens, Panepistimiopolis, Zografos, Athens
15784, Greece}
\author{K. Vetas}
\affiliation{Department of Mathematics, University of the Aegean, Karlovassi, 83200
Samos, Greece}
\affiliation{Department of Mathematics, Statistics and Physics, College of Arts and
Sciences, Qatar University, P. O. Box 2713, Doha, Qatar}

\begin{abstract}

We consider the asymptotic behavior of the solutions of a nonlinear
Schr\"odinger (NLS) model incorporating linear and nonlinear gain/loss.
First, we describe analytically the dynamical regimes (depending on
the gain/loss strengths), for finite-time collapse, decay, and global
existence of solutions in the dynamics. Then, for all the above parametric
regimes, we use direct numerical simulations to study the
dynamics corresponding to algebraically decaying initial data.
We identify crucial differences between the dynamics of vanishing
initial conditions, and those converging to a finite constant background:
in the former (latter) case we find strong (weak) collapse or decay,
when the gain/loss parameters are selected from the relevant regimes.
One of our main results, is that in all the above regimes, non-vanishing initial
data transition through spatiotemporal, algebraically decaying waveforms. While the
system is nonintegrable, the evolution
of these waveforms is reminiscent to the evolution of
the Peregrine rogue wave of the integrable NLS limit. The parametric range of
gain and loss for which this phenomenology persists is also touched upon. 

\end{abstract}

\maketitle

\section{Introduction}

In the present paper, we investigate the dynamics of
the following nonlinear Schr{\"o}dinger (NLS) model:
\begin{eqnarray}
\label{eq1}
\!\!\!\!\!\!\!\!\!
{\rm i}\partial_t u + \frac{1}{2} \partial_x^2 u + |u|^2u
&=&{\rm i}\gamma u+{\rm i}\delta|u|^2u,
\quad x \in \mathbb{R}, \quad t\in [0,T_{\mathrm{max}}).
\end{eqnarray}
Here, $u(x,t)$ is the unknown complex field, subscripts denote partial derivatives,
and
constants $\gamma,~\delta \in \mathbb{R}$ describe gain or loss (see below).
We supplement Eq.~(\ref{eq1}) with the periodic boundary conditions for $u$:
\begin{equation}\label{eq2}	
u(x+2L,t)=u(x,t),
\end{equation}
for given $L>0$, and with the initial condition
\begin{equation}\label{eq3}
u (x,0)=u_0(x),
\end{equation}
also satisfying the periodicity conditions~(\ref{eq2}).

Equation~(\ref{eq1}) is one of the most fundamental NLS models
incorporating linear and nonlinear gain/loss effects.
In particular, the parameter $\gamma$ describes linear loss ($\gamma<0$) [or gain
($\gamma>0$)],
while $\delta$ describes
nonlinear gain ($\delta>0$) [or loss ($\delta<0$)].
The presence of these
effects is physically relevant in the context of nonlinear optics
\cite{KodHas87, Agra1, Agra2, akbook, Gagnon}: in this context, $\gamma$ describes a
linear absorption ($\gamma<0$) [or linear amplification ($\gamma>0$)], while $\delta$
stands for nonlinear amplification ($\delta>0$) [or gain saturation-two photon
absorption ($\delta<0$)]. Generally,
for physically relevant settings, such terms play a crucial role in the stability of
solitons
potentially rendering them attractors for the dynamics. Note that the integrable limit
of $\gamma=\delta=0$, i.e., the conservative NLS equation, possesses soliton
solutions;
in the case of the self-focusing nonlinearity considered herein, there exist bright
soliton
solutions that decay to zero at infinity.

Apart from bright solitons, and again in the integrable version of Eq.~(\ref{eq1}),
there exists still another important class of solutions:
a rational solution, namely the Peregrine rogue wave (PRW), as well as
space- or time-periodic solutions, referred to as the Akhmediev and Kuznetsov-Ma
(KMb) breathers, respectively. These solutions have emerged from the seminal works of
Peregrine~\cite{H_Peregrine}, Kuznetsov~\cite{kuz}, Ma~\cite{ma}, and
Akhmediev~\cite{akh},
as well as Dysthe and Trulsen~\cite{dt}. In recent years, the aforementioned wave
structures are
receiving increasing attention, due to their argued potential relevance towards the
mathematical description of extreme events and rogue waves \cite{k2a,k2b,k2c,k2d}.
The interest in these solutions is justified by many relevant
experimental observations in various physical contexts, ranging from hydrodynamics
\cite{hydro,hydro2,hydro3} and nonlinear optics \cite{opt1,opt2,opt3,opt4,opt5,laser},
to superfluidity \cite{He} and plasmas \cite{plasma}.

An important question concerns the robustness of rational solutions in the presence
of integrable and non-integrable perturbations. In selected special cases
it was found~\cite{devine, NR4Anki, NR5Wang, NRbor6, NRbor5a}, that  rational solutions
can still be identified,  in Hirota-type variants
of the original NLS equation (see also Ref.~\cite{calinibook} for relevant work in
higher-order NLS models).
However, in broader classes of non-integrable perturbed NLS models, this
remains a significant open problem.

Another approach towards exploring the robustness of these excitations relies on
the connection with the KMb state. The
latter, is periodic in the evolution variable and, in the limit of infinite period,
reduces to the PRW. The dynamical robustness of KMb against dispersive or diffusive,
higher-order perturbations was studied by means of a perturbed inverse scattering
transform approach  \cite{KalimerisGarnier}. In the latter work, it was shown that KMb
is rather robust
against dispersive perturbations, but it is strongly affected by dissipative ones. In
another
related recent work \cite{KMS}, a Floquet stability analysis of the KMb was performed
in the integrable NLS, attempting --via the approach to the infinite period limit-- to
shed light to the stability features of the PRW state.

More relevant to our concerns in the present work, are recent contributions
studying the role of driving and linear loss. %
Reference \cite{onorato1}, for instance (see also references therein), investigates the
effect
of wind and dissipation on the nonlinear stages of the modulational instability
(MI) in the context of water waves \cite{ZO}. Further extensions considering the
case of strong wind (whose effect is of the same order as the wave
steepness), are given in Ref.~\cite{brunetti}. This approach, allows for the
construction of approximate
rational solutions for the considered model. The effects of slow/strong linear forcing
in the statistics of
extreme events, were also studied \cite{EPeli}. It was highlighted therein, that when
the forcing
is short in time and effectively strong, it facilitates
the occurrence of extreme wave conditions.
At this point, it is important to note that the MI of background plane waves
(which host the rational  solutions), may lead to homoclinic-type solutions,
which can also considered as candidates for rogue waves \cite{calini2012}.
Such phenomena have been observed and analyzed in a variety of extended NLS models,
that incorporate higher-order dissipative or dispersive effects (pertinent studies
also include the Dysthe model \cite{NR11}).

In this work, we aim to examine a rather different question,
related to the PRW. In particular, its algebraic spatiotemporal
localization, motivates us to study the dynamics of Eq.~(\ref{eq1}) for initial
data with an algebraic decay rate. Our plan is as follows. First, we
identify analytically the regimes defined by the gain/loss parameters
where the solutions may exhibit finite time collapse, vanishing decay, or
nontrivial (non-decaying) dynamics. The latter are captured by an  attracting set for
all initial conditions in the infinite dimensional space.
Then, we proceed to examine, by means of direct numerical simulations,
the evolution of algebraically decaying initial data in the prescribed
dynamical regimes. Our main findings
are summarized below.

To begin, crucial differences are detected between the dynamics
corresponding to asymptotically vanishing or non-vanishing initial data.
For vanishing initial conditions, we observe the strong
collapse/decay of the resulting waveforms \cite{Babis}.
The collapse is manifested by the local increase of amplitude
which grows towards the singularity,
while decay is manifested by a flattening, leading
towards convergence to the trivial state.
On the other hand, nonvanishing initial conditions exhibit a weak
type of collapse/decay. In weak collapse, as time progresses,
a fraction of the wave approaches the singularity with long tails left behind. In our
case,
these tails consist of a multi-hump, non-collapsing wave-train of solitonic
structures,
occupying a finite spatial interval of increasing length as time increases.
Collapse occurs through time-oscillations of increasing amplitude of the central
localized waveform. In the decay regime, a multi-peak wave pattern forms, and
eventually
the amplitude decays.

The second finding is, that in all the above gain/loss parametric regimes,
spatiotemporal
algebraically decaying structures emerge transiently. The
evolution of these structures is reminiscent, for a finite interval,
of that of the PRW of the integrable NLS limit. It should be remarked that the
excitation of PRW-type dynamics by the interaction between a continuous wave and a
localized (single peak) perturbation pulse, was first reported in \cite{Yang1, Yang2},
for a higher-order, non-integrable NLS (including the integrable Hirota equation
\cite{Hirota}, as a particular limit).

In the case of Eq.~(\ref{eq1}), to examine the spatiotemporal growth and decay rates of
these extreme events, we implement
effective fitting arguments, as in Ref.~\cite{P2}, comparing the profiles of the
numerical
solutions of the problem (\ref{eq1})-(\ref{eq3}), with the evolving profiles of the
PRW
of the integrable NLS limit. Both in the collapse and decay regimes, the appearance of
these structures is associated with the weak
collapse/decay of the solutions.

In the limit-set regime $\gamma>0,\,\delta<0$, we observe the generation of extreme
events in distinct dynamical regimes: the suitable localization of the initial data
results in the emergence of a PRW-type structure
at an early stage, seeding subsequently the MI.
The monotonicity properties of the evolution of suitable functionals, combined with
the continuous dependence on the initial data, is complemented by numerical findings
on the evolution of the centered density, in partially rationalizing the
observations of the PRW-like
structures.

We remark that, in all the gain/loss parametric regimes,
the emergence of the above structures persists up to numerically detected critical
values
$\gamma_{\mathrm{crit}},\delta_{\mathrm{crit}}\sim O(10^{-1})$, at most.  These critical values define threshold sub-domains visualized as orthogonal or trapezoidal regions, within the relevant parametric quadrants (see Fig.~\ref{fig1}), supporting the emergence  of PRW-type events. The existence of the above
critical values, shows
that the appearance of extreme events may occur when the NLS model (\ref{eq1}) is
sufficiently close
to its integrable limit (but not for arbitrary ranges of gain and loss). 

The summary of the above observations is that the dynamics of Eq.~(\ref{eq1})
serves as a case example suggesting that a corroboration of the appropriate
spatial localization of the initial data, with suitable energy dissipation/source
effects,
could be a potential mechanism for the emergence of extreme events in nonintegrable
nonlinear dispersive systems.

The paper is structured as follows. In Section~II, we present
an overview of analytical considerations on the problem (\ref{eq1})-(\ref{eq3}).
In Section~III, we report the results of our numerical simulations. Finally,
in Section~IV, we summarize and discuss the implications of our results with an eye
towards future work.

\section{Analytical considerations}
\label{SECI}
In this section, we describe analytically the global dynamical regimes driven by the
gain/loss strengths
$\gamma$ and $\delta$. It will be also important for analyzing the dynamics, to comment
on the modulation
instability (MI) of continuous waves (cw). Since our analysis is also valid for weak
solutions, it is
important to recall the following local-in-time existence and uniqueness result, whose
proof is
now considered as standard \cite{Caz03,Kato0}:

 \begin{theorem}\label{thmloc}
 	Let $u_0\in H^k_{\mathrm{per}}(\mathcal{Q})$ for any integer $k\geq 0$, and
 $\gamma,\delta\in\mathbb{R}.$
 	Then there exists $T_{\mathrm{max}}>0$, such that the problem
 	(\ref{eq1})-(\ref{eq3}), has a unique
 	solution
 	\begin{equation*}
 	u\in C([0,T_{\mathrm{max}}), H^k_{\mathrm{per}}(\mathcal{Q})) \quad\mbox{and}\quad
 u_t\in
 	C([0,T_{\mathrm{max}}), H^{k-2}_{\mathrm{per}}(\mathcal{Q})).
 	\end{equation*}
 	Moreover, the solution $u\in H^k_{\mathrm{per}}(\mathcal{Q})$ depends continuously
 on the initial data $u_0\in H^k_{\mathrm{per}}(\mathcal{Q})$, i.e., the solution
 operator
 	\begin{eqnarray}
 	\label{DSa}
 	\mathcal{S}(t): H^k_{\mathrm{per}}(\mathcal{Q})&\mapsto&
 H^k_{\mathrm{per}}(\mathcal{Q}),\;\;\;\;t\in[0,T_{\mathrm{max}}),\\
 	u_0&\mapsto&\mathcal{S}(t)u_0=u,\nonumber
 	\end{eqnarray}	
 	is continuous.
 \end{theorem}
Here, $H^k_{\mathrm{per}}(\mathcal{Q})$ denote the Sobolev spaces of $2L$- periodic
functions on the fundamental interval $\mathcal{Q}=[-L,L]$. We recall for the sake of
completeness their definition:
\begin{eqnarray}
\label{defSob}
H^k_{\mathrm{per}}(\mathcal{Q})&=&\{u:\mathcal{Q}\rightarrow \mathbb{C},\;\;
u\;\mbox{and}\; \frac{\partial^ju}{\partial x^j}\in L^2(\mathcal{Q}),\;\;
j=1,2,...,k;\nonumber\\
&&u(x),\;\;\mbox{and}\;\;\frac{\partial^ju}{\partial x^j}(x)\;\mbox{for
$j=1,2,...,k-1$, are $2L$-periodic}
\}.
\end{eqnarray}
We also recall that $H^{-s}_{\mathrm{per}}(\mathcal{Q})$, for $s>0$, stands for the
dual space of $H^s_{\mathrm{per}}(\mathcal{Q})$, i.e., the space of bounded linear
functionals on $H^s_{\mathrm{per}}(\mathcal{Q})$.
\paragraph{Gain-loss dynamical regimes.} The description of the gain/loss dynamical
regimes
is based on energy arguments. These arguments consider the evolution of an
energy functional, namely:
\begin{eqnarray}
\label{eq4a}
M(t)=\frac{{{e}^{-2\gamma t}}}{2L }\int_{-L}^{L}|u(x,t)|^2dx,
\end{eqnarray}
stemming from the ``power balance'' equation:
\begin{eqnarray}
	\label{OL2ge}
	\frac{d}{dt}\int_{-L}^{L}|u|^2dx=2\gamma\int_{-L}^{L}|u|^2dx
	+2\delta\int_{-L}^{L}|u|^4dx,
	\end{eqnarray}
satisfied by any sufficiently smooth, local-in-time solution of Theorem~\ref{thmloc}.
Then, applying Theorem~\ref{thmloc} for $k\geq 2$, and
employing the arguments of Refs.~\cite{PartI,P6}, we can prove the following Theorem:
\begin{theorem}
	\label{T1a}
	\begin{enumerate}
\item (Collapse in finite time).
  For $u_0\in  H^k_{\mathrm{per}}(\mathcal{Q})$, $k\geq 2$, let
  $\mathcal{S}(t)u_0=u\in C([0,T_{\mathrm{max}}), H^k_{\mathrm{per}}(\mathcal{Q}))$
  be the local- in- time solution of the problem (\ref{eq1})-(\ref{eq3}), with
  $[0,T_{\mathrm{max}})$ be its maximal interval of existence. Assume that the
  parameter $\delta>0$ and that the initial condition  $u_0(x)$ is such that
  $M(0)>0$.
      Then, $T_{\mathrm{max}}$ is finite, in the following cases:
	\begin{eqnarray}
	\label{a1a}
	&&(i)\;\;\;\;\;\;\;\;\;\;\;{{T}_{\mathrm{max}}}\le \frac{1}{2\gamma }\log \left[
1+\frac{\gamma }{\delta M\left( 0 \right)}
\right]:=\hat{T}_{\mathrm{max}}[\gamma,\delta,M(0)],\\
	\label{a2a}
	&&\mbox{for}\;\;\gamma\neq 0,\;\;\mbox{and}\;\;\
	\gamma>\gamma^*:=-\delta M\left( 0 \right).\\
	\label{b1a}
	&&(ii)\;\;\;\;\;\;\;\;\;\;\;{{T}_{\mathrm{max}}}\le \frac{1}{2\delta M\left( 0
\right)}:=\tilde{T}_{\mathrm{max}}[\delta,M(0)],\;\;\mbox{for}\;\;\gamma=0.
	\end{eqnarray}
\item (Global in time existence in $L^2$-Existence of a limit set): Let the initial
    condition $u_0$ be as above, and assume that $\gamma>0$, $\delta<0$. Then, the
    solution of the problem (\ref{eq1})-(\ref{eq3}),
$u\in C([0,\infty), L^2(\mathcal{Q}))$ and $u_t\in
C([0,\infty), H^{-1}_{\mathrm{per}}(\mathcal{Q}))$. Furthermore, an extended
dynamical system
\begin{eqnarray}
\label{wds1}
\varphi(t, u_0): H^k_{\mathrm{per}}(\mathcal{Q}))\rightarrow
L^2(\mathcal{Q}),~~\varphi(t, u_0)=u,
\end{eqnarray}
 is defined, whose orbits are bounded for all $t\in [0,\infty)$. For any bounded set
 $\mathcal{B}\in H^k_{\mathrm{per}}(\mathcal{Q})$, $k\geq 2$,  its $\omega$-limit set
 $\omega(\mathcal{B})$, is weakly compact in $L^2(\mathcal{Q})$,
and compact in  $H^{-1}_{\mathrm{per}}(\mathcal{Q})$.
\item (Full decay regime): Let $\gamma, \delta<0$. Then
    $\lim_{t\rightarrow\infty}||u(\cdot, t)||_{L^2}=0$.
\item (Behavior of the analytical upper bound for the collapse times):
    $\lim_{\gamma\rightarrow\gamma^*}\hat{T}_{\mathrm{max}}[\gamma,\delta,M(0)]=+\infty$,
    and $\lim_{\gamma\rightarrow
    0}\hat{T}_{\mathrm{max}}[\gamma,\delta,M(0)]=\tilde{T}_{\mathrm{max}}[\delta,M(0)]$.
\end{enumerate}
\end{theorem}
For our computations below it is relevant to mention that, in
the collapse regime, the functional $M(t)$ is strictly monotonically increasing, i.e.,
\begin{eqnarray}
\label{monot}
0<M(0)<M(t),\;\;\mbox{for all}\;\;t\in [0,T_{\mathrm{max}}).
\end{eqnarray}
On the other hand, in the decay regime, the average power
$P_{a}[u(t)]:=\frac{1}{2L}\int_{-L}^{L}|u(x,t)|^2dx$
is strictly monotonically decreasing, namely:
\begin{eqnarray}
\label{monotd}
0<P_{a}[u(t)]<P_{a}[u(0)],\;\;\mbox{for all}\;\;t\in [0,\infty).
\end{eqnarray}
It is also worth noticing that from the definitions of the analytical upper bounds for
collapse times, (\ref{a1a}) and (\ref{a2a}), it follows that
$\hat{T}_{\mathrm{max}}[\gamma,\delta,M(0)]$
is finite for $\gamma>\gamma^*$. Then, the limit
$\lim_{\gamma\rightarrow\gamma^*}\hat{T}_{\mathrm{max}}[\gamma,\delta,M(0)]=+\infty$,
suggests that if $\delta>0$, then the critical 
value $\gamma^*$ acts as a critical point separating the two dynamical behaviors:
collapse in-finite-time for $\gamma>\gamma^*$, and global existence for
$\gamma<\gamma^*$.

\begin{figure}[tbp]
	\begin{center}
		\begin{tabular}{c}
			\includegraphics[scale=0.21]{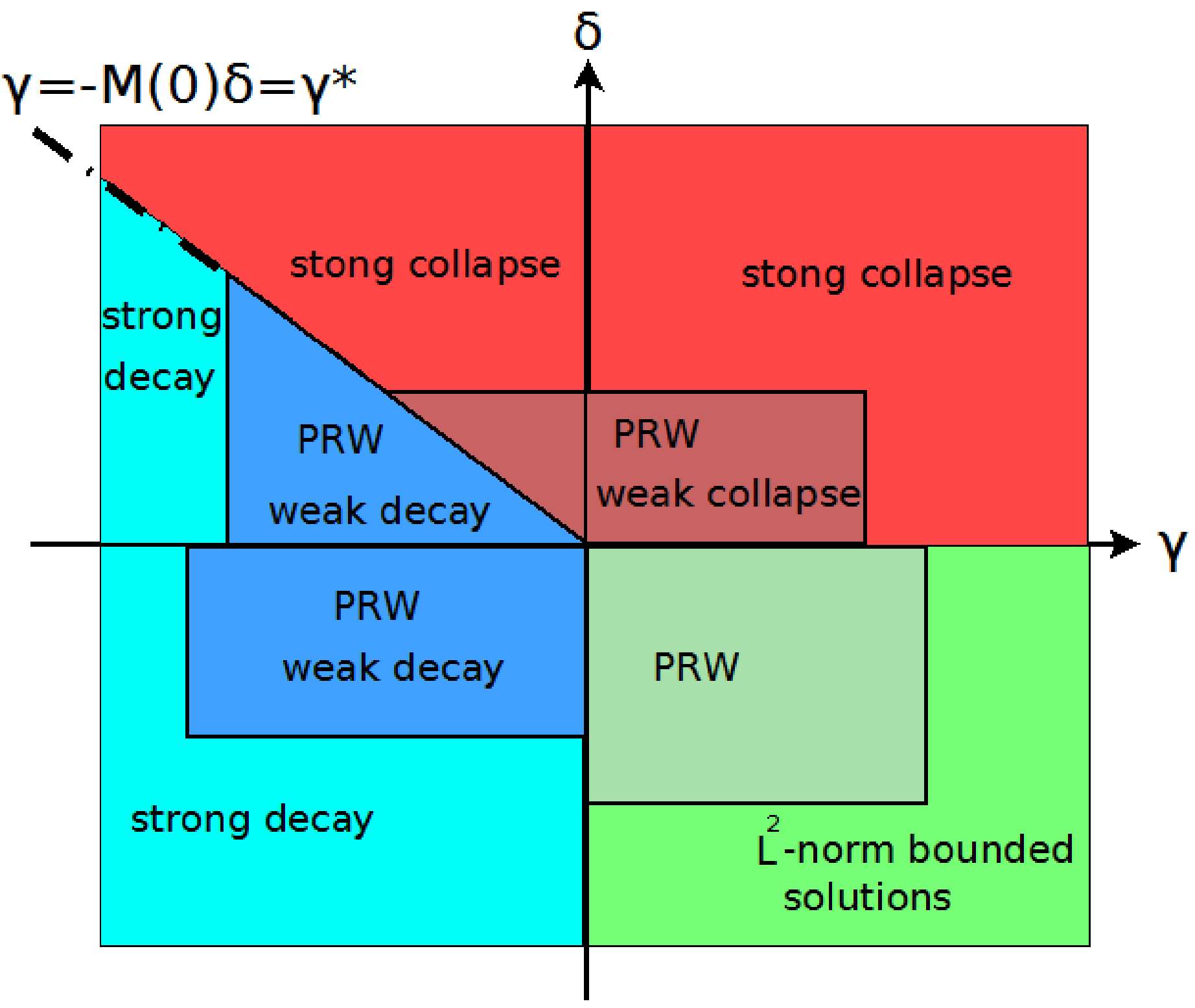}
		\end{tabular}
		\caption{(Color Online) The global asymptotic behavior of solutions
in the $\gamma$-$\delta$ plane.
		The 4th quadrant is characterized by the existence of a limit set in $L^2$. Sub-domains sustaining PRW-type waveforms are shown in each quadrant, defined by threshold values $\gamma_{\mathrm{crit}},\delta_{\mathrm{crit}}\sim O(10^{-1})$, at most (cartoon). }
		\label{fig1}
	\end{center}
\end{figure}

\paragraph{Summary on the global dynamics.}
According to our analytical arguments given above, we can depict
schematically --cf. Fig.~\ref{fig1}--
the dynamical scenarios for the problem (\ref{eq1})-(\ref{eq3})
for the different regimes of the gain/loss parameters.

The limit set is the trivial one, i.e., $\omega(\mathcal{B})=\left\{0\right\}$,
in the 2nd quadrant, for $\gamma<\gamma^*$ and $\delta>0$, as well as in the 3rd
quadrant,
for $\gamma,\,\delta<0$. We also note that any solution of Eq.~(\ref{eq1}) in
the form of a continuous wave (cw) is always unstable:
indeed, as discussed in Ref.~\cite{P6}, cw waves
either decay or grow in the 1st, 2nd and 3d quadrants of Fig.~\ref{fig1}, while for
cw amplitude $A=-\gamma/\delta$
in the 4th quadrant, it is prone to the modulational instability (MI). Therefore,
such cw solutions could not occur generically, as limit-points $\omega(u_0)$,
included in the attracting set $\omega(\mathcal{B})$.
Nevertheless, a broad class of initial conditions of the form:
\begin{eqnarray}
\label{inpw1}
u_0(x)=A(x)\exp\left(-\mathrm{i}\frac{K\pi x}{L}\right),\quad K>0,
\end{eqnarray}
of amplitude $A(x)$ such that $A(x) \rightarrow h_0 \ne 0$ as $x \rightarrow \pm
\infty$,
and wavenumber $K \ge 0$ may lead to plane wave evolution
and the manifestation of the corresponding MI effects \cite{P6}.
We finally note that if $A(x)=$const., then the upper bounds of the collapse times are
sharp \cite{PartI}.

\section{Numerical Results}
\subsection{Background.}
Selecting parameters $\gamma$ and $\delta$ within the dynamical
regimes prescribed in Sec.~\ref{SECI}, we will proceed by numerically studying
the dynamics corresponding to algebraically decaying initial data.
This is the principal focal point of the present contribution, motivated
by our aim
to assess the relevance of PRW-type states
in models involving driving or gain and damping.
More precisely, we will examine
the evolution of initial conditions of the following form:
\begin{eqnarray}
\label{ic1}
	u_0(x)=h_0+\frac{c_1}{c_2+c_3x^2}, \quad c_i>0, \quad i=1,2,3.
\end{eqnarray}
The initial condition (\ref{ic1}) can be partitioned in two cases:
(a) $h_0=0$ and (b) $h_0>0$, corresponding to a quadratic decay to a vanishing
or to a finite background, respectively.
The second of the above initial conditions is, in fact, a PRW profile at $t=0$:
\begin{eqnarray}
\label{ic2}
	u_0(x)=u_{\mbox{\tiny
PS}}(x,t;t_0;P_0)|_{t=0}:=\sqrt{P_0}\left\{1-\frac{4\left[1+2\mathrm{i}P_0(t-t_0)\right]}{1+4P_0x^2+4P_0^2(t-t_0)^2}\right\}\mathrm{e}^{\mathrm{i}P_0(t-t_0)}|_{t=0},
\end{eqnarray}
is the PRW solution (translated at $t=t_0$) of the integrable NLS,
corresponding to $\gamma=\delta=0$.


The PRW is the prototypical solitary waveform, possessing power-law (quadratic)
rather than the more standard exponential spatiotemporal localization.
For this reason, our aim is to use the PRW profile~(\ref{ic2}),
and employ fitting arguments in the spirit of Ref.~\cite{P2}, to examine the
spatiotemporal growth or decay rate of the numerical solutions of
(\ref{eq1})-(\ref{eq3}).
In this way, our scope is to quantify, starting in the present stage with numerical
investigations, the potential robustness of PRW in the presence of perturbations,
i.e., for $\gamma,~\delta \ne 0$.

It should also be mentioned that the initial conditions (\ref{ic1}) and (\ref{ic2})
can be considered as elements of weighted $L^2$, or Sobolev spaces, defined as
follows.
Consider the weight function $w(x)= 1+|x|$. Then, the initial conditions
(\ref{ic1}) or (\ref{ic2}) are such that:
\begin{eqnarray}
\label{defXn}
u_0\in
X_{w}=\left\{u:\mathcal{Q}\rightarrow\mathbb{C}\;\;:\;\;\int_{\mathcal{Q}}w^2(x)|u-h_0|^2dx<\infty
\right\}.
\end{eqnarray}
In particular, $u_0\in X^1_{w}(\mathcal{Q})\cap H^2(\mathcal{Q})$, where the weighted
Sobolev spaces $X^m_{w}$ are defined as:
\begin{eqnarray*}
	X^m_{w}=\left\{u:\mathcal{Q}\rightarrow\mathbb{C}\mbox{:}\; u\in
X_w\;\;\mbox{and}\;\; \sum_{j=0}^m
	\int_{\mathcal{Q}}w^2(x)|\partial^j_xu|^2dx<\infty\right\}.
\end{eqnarray*}

When $L$ is finite but sufficiently large, or $L\rightarrow\infty$, an initial
condition
$u_0\in X^1_{w}(\mathcal{Q})\cap H^2(\mathcal{Q})$ and its derivative converge with,
at
least, a quadratic decay-rate to $h_0$ and zero, respectively, and has essentially
a localization width $\ll 2L$.
Note also that, as a consequence of Theorem~\ref{thmloc}, the continuous dependence on
the initial data in the above space can be stated as:
\begin{eqnarray}
\label{condep}
||\mathcal{S}(t)u_0-\mathcal{S}(t)u_1||_{X^1_{w}\cap H^2}\leq
K(L,T_0)||u_0-u_1||_{X^1_{w}\cap H^2},\quad K(L,T_0)>0, \quad T_0\in
[0,T_{\mathrm{max}}),
\end{eqnarray}
which will be useful for the analysis of the numerical results: actually we
will consider the space  $X^1_{w}(\mathcal{Q})\cap H^2(\mathcal{Q})$ as
a potential domain of attraction for PRW-reminiscent waveforms.

Since  numerical simulations consider periodic boundary conditions, we should point out
that  they are satisfied by the initial conditions (\ref{ic1}) and (\ref{ic2}) (or
generically, by  elements of $X_w$), only asymptotically as $L\rightarrow\infty$.
However, the  largest error in these asymptotics is of order O$(L^{-2})$. Since the
smallest value of $L$ used in our numerical experiments is $L=50$, the error is of
order O$(10^{-4})$ or less, and
we have confirmed that it has negligible effects in the dynamics.

The numerical experiments have been performed by using a high accuracy pseudo-spectral
method, incorporating an embedded error estimator that makes it possible to efficiently
determine appropriate step sizes. Regarding the simulations on collapse, we denote as
collapse time
the time where the numerical scheme detects a singularity in the dynamics.
The numerical singularity is detected when the time step is becoming of order
$<10^{-12}$.
\begin{figure}[tph!]
	\hspace{-0.2cm}
	\includegraphics[scale=0.207]{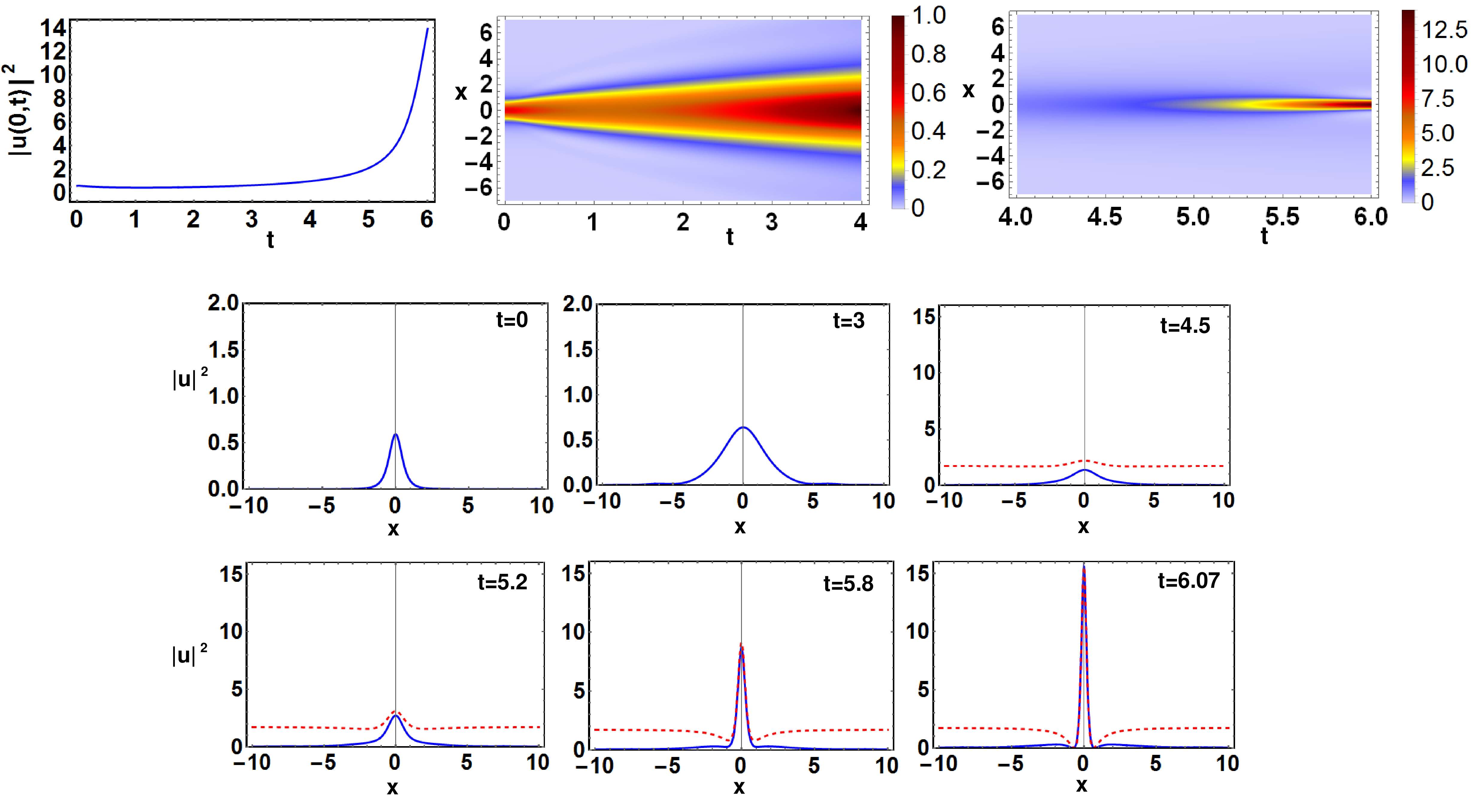}
	\caption{(Color Online) 
		Top row: The left panel shows the evolution of the density of the center,
		$|u(0,t)|^2$,
		for the initial condition (\ref{ic1}), with vanishing background $h_0=0$,
		and $c_1=1$, $c_2=1.3$, $c_3=2$. Parameters $\gamma=0.2$, $\delta=0.01$, $L=50$. The middle and right panels show contour plots for the
		evolution of the density $|u(x,t)|^2$,
		for the same evolution, as in the left panel.
		Middle (right) panel corresponds to $t\in [0,4]$
		($t\in [4,6]$).
		Bottom rows: snapshots of the density $|u(x,t)|^2$
		[solid (blue) curves] at different times, for the above evolution.
		The evolution of $|u(x,t)|^2$ is compared to that
		of the PRW $u_{\mbox{\tiny PS}}(x,t;6.07;1.73)$ [dashed (red) curve]. It can
		be seen that, modulo its absence of background, the central
		portion of the solitary wave closely resembles, in the vicinity
		of its peak amplitude time, the PRW structure.}
	\label{figure2}
\end{figure}
\subsection{Collapse regime.} We start our presentation with the case
$\gamma,\,\delta>0$ (1st-quadrant in the $\gamma\delta$-plane of Fig.~\ref{fig1}),
corresponding to the full collapse regime.
%
\begin{figure}[tbh!]
	\hspace{-0.5cm}\includegraphics[scale=0.21]{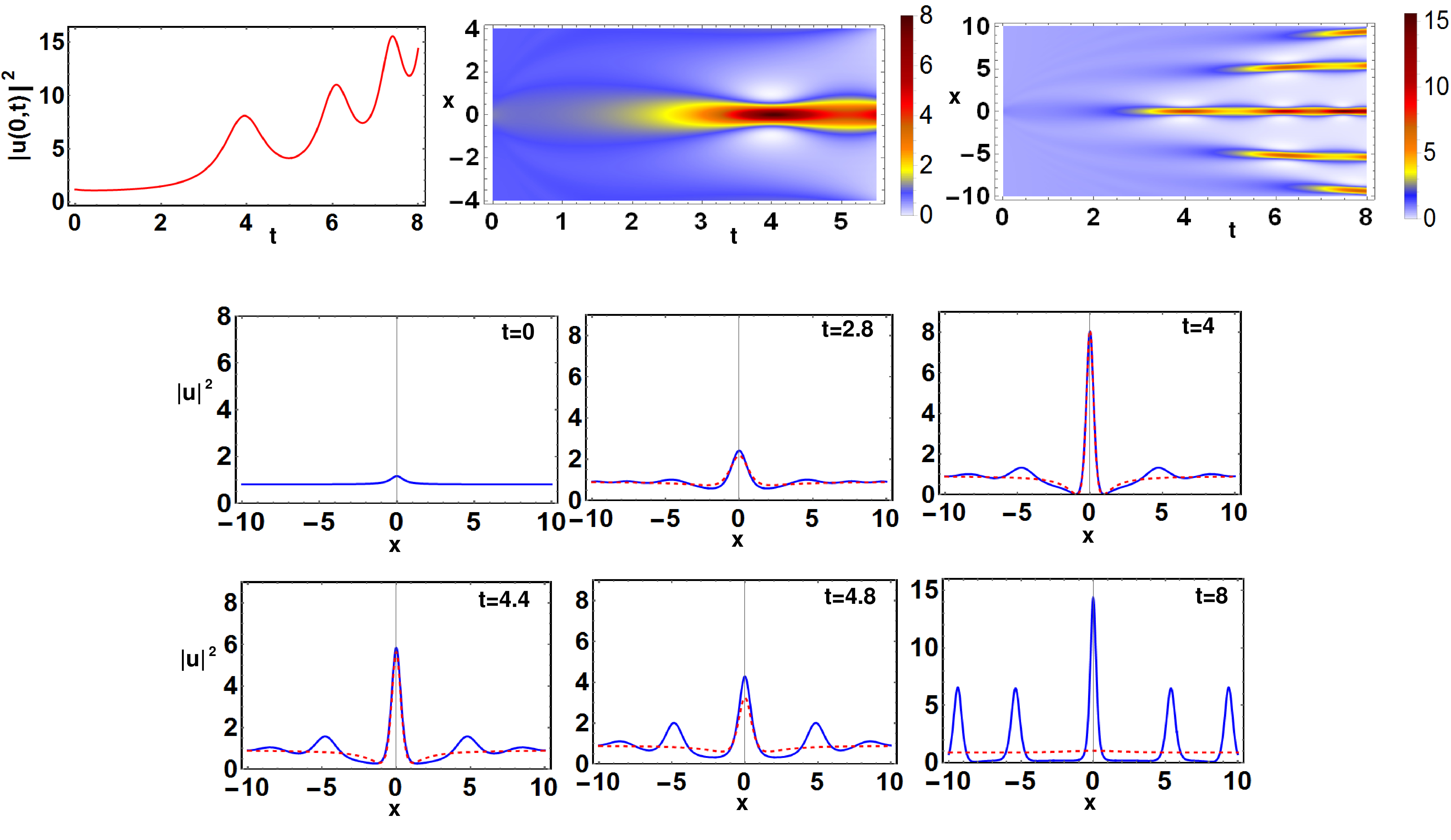}
	\caption{(Color Online) 
	Top row: The left panel shows the evolution of the density of the center, $|u(0,t)|^2$, for the initial condition
	(\ref{ic1}), with non-vanishing background $h_0=0.9$, and $c_1=0.18$, $c_2=1$, $c_3=3.6$. Parameters $\gamma=0.01$, $\delta=0.01$, $L=50$. The middle and right panels show contour plots for the
	evolution of the density  $|u(x,t)|^2$, for the same evolution, as in the left panel.
Left (right) panel
	corresponds to $t\in [0,6]$
	($t\in [0,8]$). Bottom rows: snapshots of the density $|u(x,t)|^2$ [solid (blue)
curve] at different times, for the above evolution.
The evolution of $|u(x,t)|^2$ is compared to that
of the PRW $u_{\mbox{\tiny PS}}(x,t;4;0.9)$ [dashed (red) curve]. }
	\label{figure3}
\end{figure}
The left panel of the top row of  Fig.~\ref{figure2}, shows the evolution of
the density at the center, $|u(0,t)|^2$, for the initial condition (\ref{ic1}) with zero-background $h_0=0$, and $c_1=1$, $c_2=1.3$, $c_3=2$.
The rest of parameters are 
$\gamma=0.2$, $\delta=0.01$, and $L=50$. 
Apart from a slight decrease of $|u(0,t)|^2$ during the early stages
of the dynamics, i.e., within the interval $[0,2]$, we observe an
almost monotonic increase, which is typical of a strong collapse \cite{Babis}:
since $|u(0,t)|^2$ increases almost monotonically, and the functional $M(t)$ is
monotonically increasing as discussed before,
most of the initial energy will be concentrated to a single localized structure
near the singularity.

Such behavior is illustrated in the middle and the left panel of the top row of Fig.~\ref{figure2}, showing contour plots
depicting the evolution of the density $|u(x,t)|^2$;
the middle (right) panel shows the evolution for $t\in[0,4]$ ($t\in[4,6]$), illustrating the
initial slight decrease (mentioned above) and the subsequent
increase of the amplitude for $t>2$. This behavior is also observed
in the snapshots of the evolution of the density [solid (blue) curve],
portrayed in the second row. The right panel of this
row reveals that as the singularity is approached, the mass becomes progressively
concentrated around the
center. In addition,  we compare the profiles of the numerical solution of
(\ref{eq1})-(\ref{eq3})
with the PRW profile of the integrable limit, $u_{\mbox{\tiny PS}}(x,t;6.07;1.73)$
[dashed (red) curve].
The value of $P_0$ (and $t_0$) is selected by simply matching the
time (and value) of the maximal density amplitude.
This comparison reveals that the time-growth rate, of the numerical solution
towards the singularity, is reminiscent of the time-growth rate of the PRW-profile.
It is also interesting to observe that the localized structure forms, towards
collapse,
a PRW-core (with a suitably adapted spatial distribution),
as shown in the last snapshot at $t=6.07$. This phenomenon will be further discussed
below.

The structure of collapse appears to be totally different
in the case of a finite background $h_0>0$. The left  panel of the top row of Fig.~\ref{figure3}
reveals that, in contrast to the case $h_0=0$, the evolution of the density
at the center, $|u(0,t)|^2$, is not monotonic. The
initial condition corresponds to $h_0=0.9$, with $c_1=0.18$, $c_2=1$, $c_3=3.6$, while the rest of parameters are $\gamma=0.01$, $\delta=0.01$ and $L=50$. 
The evolution in this case suggests a drastically different event prior to collapse,
which we can describe as follows.
Here, $|u(0,t)|^2$ exhibits a local maximum at $t\simeq 4$ and a local minimum at
$t\simeq 5$,
yet the functional $M(t)$ is strictly increasing. Hence,
the decrease of the center density is followed by a partial transfer
to emergent localized side lobes adjacent to the center,
while $M(t)$ remains increasing.
Therefore, the type of collapse should be weak, with the formation of tails
consisting of localized waveforms. 

The dynamics in this case is shown in the rest of the panels of Fig.~\ref{figure3}. There, the differences
between the present and the previous case are highlighted:
in the present case, a weak-type of collapse is observed, as explained above and shown
in the snapshot at $t=8$. Furthermore, another remarkable feature is that,
even in this case, a PRW-type wave form appears to emerge.
The comparison between the numerical solution of Eq.~(\ref{eq1}) and the PRW
$u_{\mbox{\tiny PS}}(x,t;4;0.9)$ within the time interval $[2.8,4.8]$
reveals that the centered localized structure
possesses a spatiotemporal growth/decay rate, reminiscent of that of the PRW.
This can be attributed to the continuous dependence on the initial data $u_0$.
At the same time, this intriguing observation
seems to indicate that the PRW structure is a transient precursor arising ``en route''
to collapse dynamics.  This is related to the broader scale question
about to which extent, extreme (yet finite amplitude)
events, such as the PRW, may catalyze --if present-- the transition towards collapse.
As seen in Fig.~\ref{figure3} and discussed above,
the continuous dependence on initial data (\ref{condep}), along with the monotonic
growth of $M(t)$, lead to a central peak that tends to gain energy from the finite
background.
At the same time, the dynamics of the --generic, here-- modulational instability,
recently analyzed in its nonlinear stage in the work of Ref.~\cite{BM},
suggests the division of the spatial domain into three regions. These are a far left
field
and a far right field, in which the solution is approximately equal to its
initial value, as well as a central region in which the solution has
oscillatory behavior with respect to the spatial variable $x$.
While the situation here is complicated due to the presence
of the gain and loss, a similar division of the domain
is portrayed --cf. top right contour plot of Fig.~\ref{figure3}.
Nevertheless, we observe time-oscillations of increasing amplitude.
In summary, the competition between the effects of core gain
and side-band growth due to MI, results in a non-monotonic
scenario for the increase of the core amplitude; this increase,
combined with the manifestation of MI, is also accompanied by the emergence of
side humps, of also increasing oscillatory amplitude. However, the growth rate of the
eventually
collapsing core, is larger than the growth rate of the developing humps.

It is important to remark, that the above type of weak collapse, and the emergence of a PRW-type wave form, persists up to small values of $\gamma>\gamma^*$ and $\delta>0$, i.e., up to critical thresholds $\gamma_{\mathrm{crit}},\delta_{\mathrm{crit}}\sim O(10^{-2})$. Beyond that critical values, the growth of  $|u(0,t)|^2$ becomes again monotonic, reminiscent to that discussed in the example of Fig.~\ref{figure2}, and a strong type of collapse is again manifested. This is a first evidence that PRW-type dynamics may emerge in $\gamma,\delta$-parametric regimes, close to the NLS integrable limit.

Concerning the behavior of the analytical upper bounds (\ref{a1a})-(\ref{b1a}) of the collapse times, when implemented for localized initial data, we remark that the type of collapse for such data [with either algebraic as (\ref{ic1}), or exponential decay as in \cite{PartI}], does not follow the ODE dynamics of the homogeneous background.  This effect is generically reflected in discrepancies between the numerical blow-up times and the upper bounds,  and analyzed in \cite[Sec.3.2, pg. 62]{PartI}.
\subsection{Decay regime.}
\begin{figure}[tbh!]
	\hspace{-1.2cm}\vspace{-0.4cm}
	\includegraphics[scale=0.22]{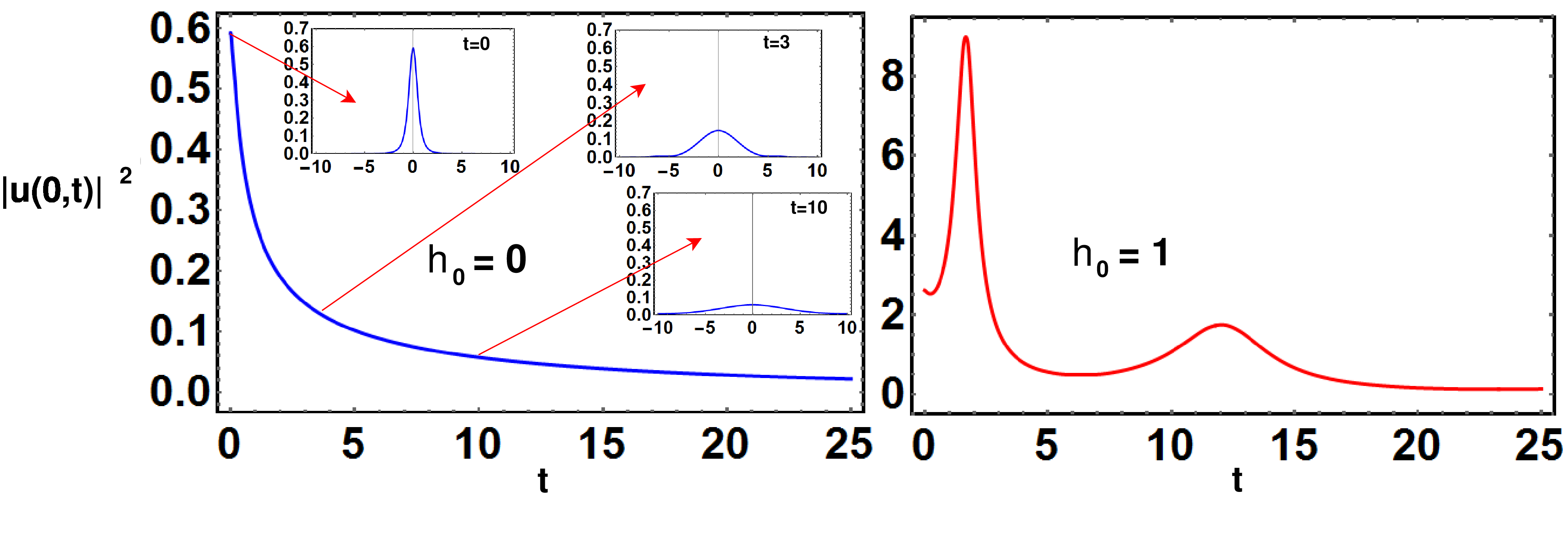}
	\caption{(Color Online) Left panel:
		monotonic decay for the initial condition (\ref{ic1}) with vanishing background
		$h_0=0$, and $c_1=1$, $c_2=1.3$, $c_3=2$;
		other parameters: $\gamma=-0.01<\gamma^*\simeq 0$, $\delta=0.01$.
		The solid (blue) curve depicts the evolution of the center density $|u(0,t)|^2$.
		Arrows are directed to insets portraying snapshots of the density $|u(x,t)|^2$
		at $t=0$ (initial condition), $t=3$ and $t=10$.
		Right panel: non-monotonic decay for an initial condition (\ref{ic1}) with finite
		background
		$h_0=1$ and  $c_1=0.8$, $c_2=1.3$ and $c_3=2$; rest of parameters $\gamma=-0.03$,
		$\delta=-0.01$
		(full decay regime), and $L=50$. Emergence of an extreme event occurs at $t=1.625$. }
	\label{figure4}
\end{figure}

We now turn to the decay regime, once again distinguishing between
initial data with $h_0=0$ and $h_0 \neq 0$.
When $h_0=0$, we observe
decay, as shown in the left panel of Fig.~\ref{figure4}.
This figure depicts the monotonic decay of the center density, $|u(0,t)|^2$, and insets
show
characteristic density snapshots at specific times.
The rest of the parameters for the initial condition are
$c_1=1$, $c_2=1.3$, $c_3=2$, as in the case of
collapse in finite time of Fig.~\ref{figure2}, for $\delta=0.01$ and
$\gamma>\gamma^*\simeq 0$.
However, while still $\delta=0.01$ as in the collapse example,
the parameter $\gamma=-0.01<\gamma^*$ leads to decay dynamics.
Clearly, this example highlights the
accuracy of the analytically defined ``separatrix'', distinguishing
collapse from decay in the 2nd quadrant in the $\gamma\delta$-plane of
Fig.~\ref{fig1}.
The decay is now explained by reversing the argument used for the strong collapse
of Fig.~\ref{figure2}: since $|u(0,t)|^2$ is monotonically decreasing, the profiles should
flatten
prior to decay, so as to follow the monotonic decrease of the averaged power
functional $P_{a}[u(t)]$  [cf. Eq.~(\ref{monotd})].
\begin{figure}[tbh!]
	\centering
	\hspace{-1.2cm}\includegraphics[scale=0.21]{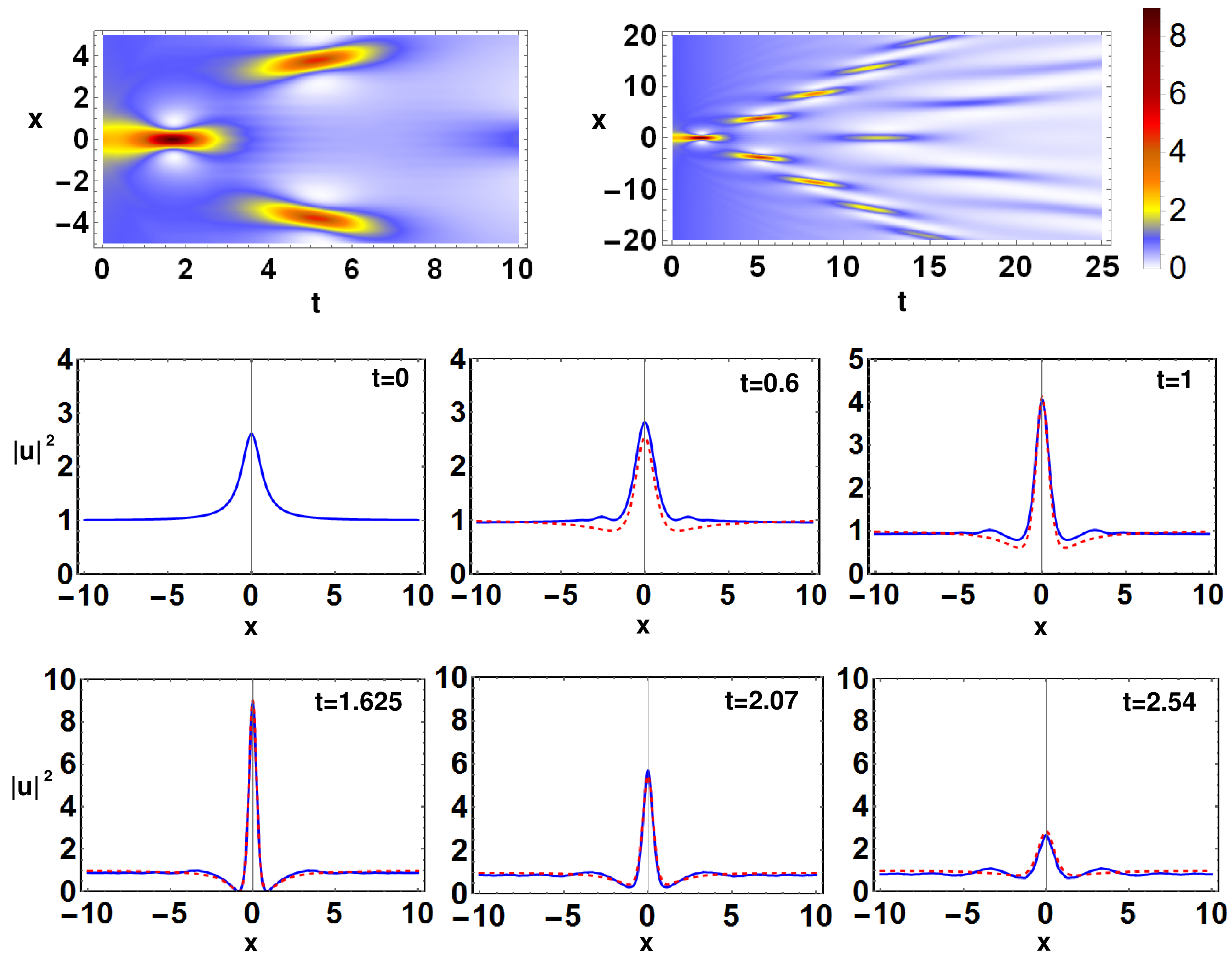}
	\caption{(Color Online)
		Top row: contour plots for the
		evolution of the density $|u(x,t)|^2$,
		for the initial condition (\ref{ic1}), with  non-vanishing background $h_0=1$,
		and $c_1=0.8$, $c_2=1.3$, $c_3=2$; other parameters: $\gamma=-0.03$,
		$\delta=-0.01$, $L=50$.
		Left (right) panel
		corresponds to $t\in [0,10]$
		($t\in [0,25]$). Bottom rows: numerically obtained density snapshots
		[solid (blue) curves] are compared to those of
		the PRW, $u_{\mbox{\tiny PS}}(x,t;1.625;1)$ [dashed (red) curves].
	}
	\label{figure5}
\end{figure}
Rather surprisingly, when the initial condition features
a finite background $h_0>0$, the emergence of a PRW-type waveform is observed in the
full decay regime
$\gamma<0,\,\delta<0$ (3rd quadrant in Fig.~\ref{fig1}). This is
particularly interesting from a physical point of view: in this regime,
Eq.~(\ref{eq1})
describes a realistic situation, namely the evolution of pulses or beams in nonlinear
optical
media featuring linear loss and two-photon absorption, which are generic dissipative
effects
in the context of optics \cite{Agra1,Agra2}.

The right panel of Fig.~\ref{figure4} depicts the non-monotonic
decay of the center density $|u(0,t)|^2$ for an initial condition (\ref{ic1}), with
$h_0=1$,
$c_1=0.8$, $c_2=1.3$ and $c_3=2$; the loss parameters are $\gamma=-0.03$ and
$\delta=-0.01$.
Evidently, an extreme event occurs at $t=1.625$ and precedes the
eventual decay dynamics of this example. This is analyzed further in Fig.~\ref{figure5}.
The first row shows contour plots for the
evolution of the density. In particular,
the left panel shows the evolution at early times, i.e., for $t\in[0,10]$, illustrating
the emergence
of an extreme event
(cf. peak of the center density in the right panel of Fig.~\ref{figure4}); the right
panel
illustrates the
evolution of the density for $t\in[0,25]$, depicting the
structure of the decaying dynamics. In the bottom rows, snapshots of
$|u(x,t)|^2$ for $t\in [0,2.54]$ are portrayed, and are compared to
the evolution of the PRW, $u_{\mbox{\tiny PS}}(x,t;1.625;1)$. The comparison clearly
supports the identification of the emerging structure around
the peak formation as one of the Peregrine family.
The centered localized waveform of the numerical solution, preserves the algebraic
spatial decay of the initial condition, but also exhibits an algebraic in
time growth/decay rate: notably, both the time-growing, and then
time-decaying, centered profiles appear to manifest
a locking to a PRW-type mode, whose maximum density is attained at $t=1.625$.
For $t\in(1.625,2.54]$, the algebraic-in-time decay rate remains
close to that of the PRW.

It is relevant to mention here, that the excitation of a PRW-type event, by localized
initial data on the top of a finite background, was observed in \cite{Yang1}, for
a  non-integrable, higher-order NLS equation. The initial condition in \cite{Yang1} is
an interaction between a Gaussian-type pulse of small amplitude and a cw. It was shown
that when the width of the initial
perturbation pulse is wide enough, a PRW can be excited.  In the same spirit, the
evolution of a rational fraction pulse, extracted from the analytical PRW-solution, was
investigated in \cite{Yang2}.
\begin{figure}[tbh!]
	\centering
	\hspace{-1.2cm}\includegraphics[scale=0.21]{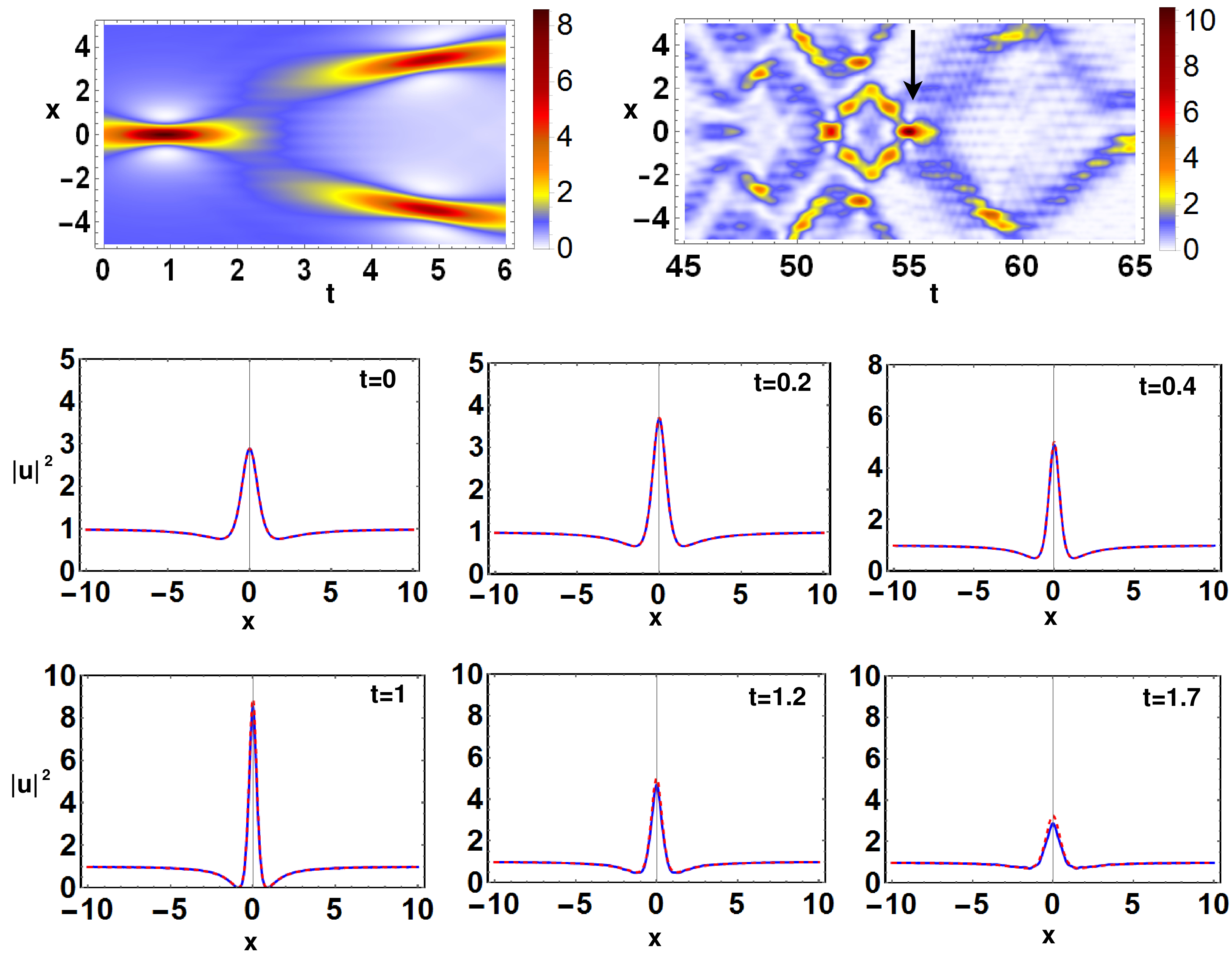}
	\caption{(Color Online)
		Top row: contour plots for the
		evolution of the density, $|u(x,t)|^2$,
		for the initial condition (\ref{ic2}), $u_0(x)=u_{\mbox{\tiny PS}}(x,0;-0.9;1)$;
		other parameters: $\gamma=0.01$, $\delta=-0.01$ (limit set regime), and
$L=50$.
		Left (right) panel
		corresponds to $t\in [0,6]$
		($t\in [45,65]$). Bottom rows: numerically obtained density snapshots
		[solid (blue) curves] are compared to those of
		the PRW, $u_{\mbox{\tiny PS}}(x,t;-0.9;1)$ [dashed (red) curves].}
	\label{figure6}
\end{figure}

In light of the results of \cite{Yang1}, here, we see the combination 
of the following effects: the proximity of the initial condition to the
  PRW, and of the model to the integrable NLS, which apparently
  drive the growth of the Peregrine phase, and --at the same time--
  the role of the $\gamma$ and $\delta$ terms in enforcing
  the monotonic decay of the functional $P_{a}[u(t)]$.
  As a result, the transient initial growth eventually gives
  way to the expected long-time decay of the model data.
  The considerations  of \cite{BM} regarding the manifestation
  of MI are still applicable in this full decay regime, explaining
  the right contour plot of Fig.~\ref{figure5}. However, the dynamics
  is eventually overwhelmed by loss events, hence the effect of MI is suppressed.

As in the case of the collapse regime, it should also be remarked that the emergence of extreme events
reminiscent of PRW, persists up to a critical threshold of
$\gamma_{\mathrm{crit}},\delta_{\mathrm{crit}}\sim O(10^{-1})$ (i.e., for a higher order of magnitude of $|\gamma|,\;|\delta|$, than in the collapse regime). However, the amplitude of the PRW-type event is decreasing, as we approach these critical values. Furthermore, beyond these critical values, the solution preserves its structural form during decay.
We qualitatively consider such thresholds as distinguishing between the
near-integrable
NLS-based regime, where gain and loss bear a more perturbative role
(even if asymptotically dominating) to the NLS dynamics
--at least in short/intermediate time scales-- and a fundamentally
  different regime. In the latter, the gain/loss parameters play
  a more drastic role, and may completely suppress the integrable
  features, including the potential formation of Peregrine-type structures.
\subsection{Limit Set Regime.} Let us finally consider
the global existence regime $\gamma>0$, $\delta<0$, which is generically associated
with the nontrivial limit set $\omega(\mathcal{B})$, discussed in Sec.~II. In this
regime,
we choose
parameter values $\gamma=0.01$, $\delta=-0.01$, and
examine the evolution of an initial condition (\ref{ic2}),
namely $u_0(x)=u_{\mbox{\tiny PS}}(x,0;-0.9;1)$.
The snapshots of the evolution of the density
in Fig.~\ref{figure6} show that the numerical solution of the nonintegrable
Eq.~(\ref{eq1}) fits very well
the PRW of the integrable limit, for $t\in [0,1.7]$.
\begin{figure}[tbp]
	\centering
	\hspace{-0.2cm}\includegraphics[scale=0.21]{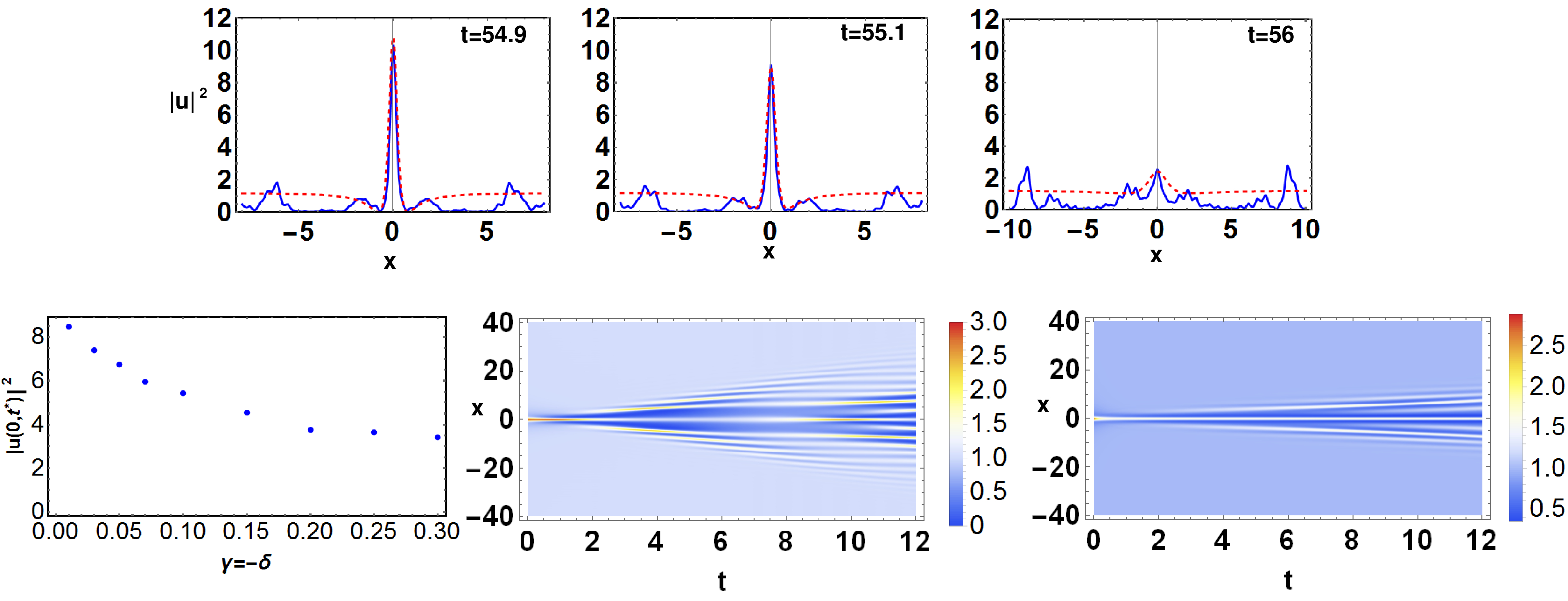}
	\caption{(Color Online) Top row: Density snapshots
		corresponding to the evolution of the Fig.~\ref{figure6}, for $t\in [54.6, 56]$.
		Density $|u(x,t)|^2$ [solid (blue) curve] is compared to the evolution of the
PRW,
		$u_{\mbox{\tiny PS}}(x,t;54.9;1.2)$ [dashed (red) curve]. Bottom row: the left panel depicts the decrease of the maximum of the density $|u(0,t^*)|^2$, of the first localized event, for increasing values of $\gamma=-\delta$. The middle and right panels, show contour plots of the evolution of the density $|u(x,t)|^2$, when $\gamma=-\delta=0.4$, and $\gamma=-\delta=1.5$, respectively. In all panels, the initial condition, and the rest of parameters, are fixed as in Fig.~\ref{figure6}. 
	}
	\label{figure7}
\end{figure}
This PRW-type structure
corresponds to the first extreme event detected in the
left panel of the first row of Fig.~\ref{figure6}.
Another interesting extreme event emerges at $t\approx 54$,
long after the MI has taken over, and led the entire domain to a highly
nonlinear, apparently chaotic phenomenology.
This event is pointed out by the (black) arrow in the right panel
of the first row of Fig.~\ref{figure6}.
Figure~\ref{figure7} shows density snapshots of
this structure for $t\in[54.6, 56]$, which are compared to the density evolution of the
PRW
of the integrable limit, $u_{\mbox{\tiny PS}}(x,t;54.9;1.2)$.
Note that the maximum of the numerical solution in this time interval, is attained
at $t\approx 54.9$. The centered localized waveform, possesses a growth/decay rate in
time,
still reminiscent of the PRW. In the vicinity of specific times (such as $t=54.9$
--where the maximum is attained-- or $t=55.1$), its profile is quite reminiscent of
that of the PRW.

Similarly to the cases of the collapse and decay regimes, there are threshold values $\gamma_{\mathrm{crit}},\delta_{\mathrm{crit}}\sim O(10^{-1})$ (i.e., of the same order of magnitude as in the decay regime), for the persistence of PRW-type dynamics. An illustrative example is portrayed in the bottom row of Figure~\ref{figure7}: the first panel shows the decrease of the maximum  density $|u(0,t^*)|^2$, of the center of the first localized event (achieved at a specific time $t^*$), as $\gamma=-\delta$ increases (i.e., as we consider increasing values along the diagonal on the 4th quadrant $\gamma>0$, $\delta<0$). The initial condition, and the rest of parameters, are fixed as in Fig.~\ref{figure6}. For increased values of $\gamma=-\delta$, the amplitude of the centered localized structure oscillates, but within moderate values, as depicted in the contour plot of the middle panel, corresponding to the case $\gamma=-\delta=0.4$.  For even larger values, the amplitude is tending to a stabilization, and the dynamics are closely reminiscent to that of \cite[Fig. 3, pg. 043902-4]{BM}, for the integrable NLS-limit. Such dynamics are portrayed in the contour plot of the right panel, corresponding to the case $\gamma=-\delta=1.5$.

Movies, related to the numerical findings illustrated in this Section,
can be found in \cite{M4a}-\cite{M9}.
\section{Discussion and Conclusions}
\begin{figure}
	\begin{center}
		\begin{tabular}{c}
			\includegraphics[scale=0.18]{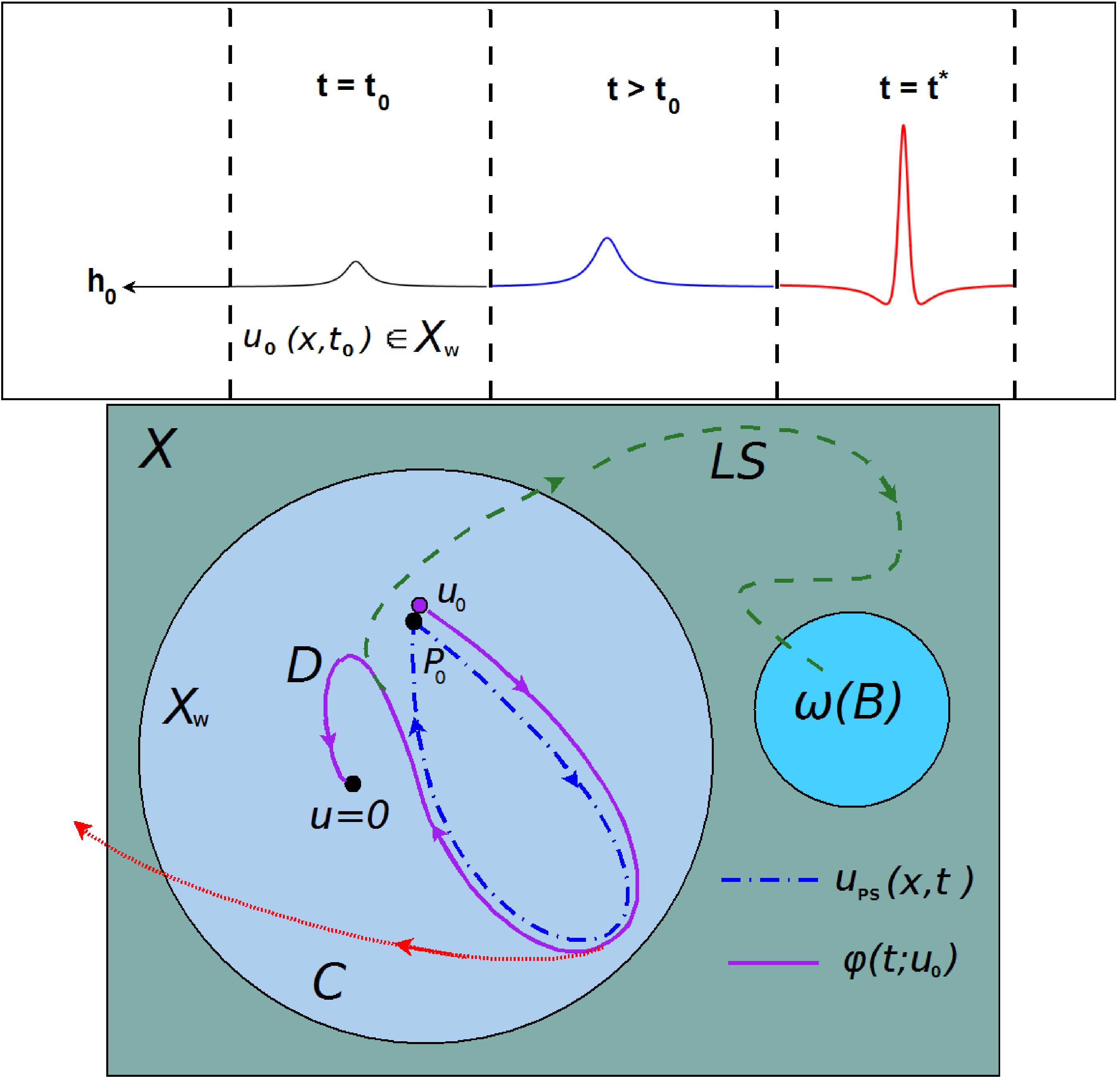}
		\end{tabular}
		\caption{(Color Online) Top panel: An initial condition $u_0(x,t_0)\in X_w$,
evolves towards  a waveform reminiscent of the PRW, achieving its maximum at $t=t^*$
(cartoon).
		Bottom panel: A potential PRW-type solution $u_{\mbox{\tiny PS}}(x,t)$ of
(\ref{eq1}), viewed as an unstable ``homoclinic loop''
		[dashed (blue) curve]], connecting $P_0$ (cartoon).}
		\label{figure8}
	\end{center}
\end{figure}
%
%
In the present work, we have considered the dynamics of
a nonlinear Schr{\"o}dinger (NLS) model,
incorporating linear and nonlinear gain/loss, and supplemented with periodic boundary
conditions.
The model finds applications in nonlinear optics, where it describes evolution of
pulses
or beams in optical media with gain and loss (e.g., linear gain or loss,
two-photon absorption, etc).

We started our exposition by outlining the parametric regimes
for the existence of a nontrivial (attracting) limit set, decay, and finite time
collapse.
Then, we examined numerically, in all the above regimes, the dynamics
stemming from initial data with
algebraic localization.
The two broad classes of initial data considered were
those decaying to zero and those converging to a finite background.
Collapse and decay were explored in both regimes and 
significant differences between them were found.
In particular, for the former case of initial conditions
the solution preserves its structural form during decay,
or develops almost strong collapse. For the latter case of initial data,
the solution decays without preserving its form, or is subject to
a weak-type of collapse. The emerging profiles at the early stage
of the evolution were analyzed by considering the combination of
the monotonicity properties of suitably defined functionals
and the modulational instability
of the background. It was thus found that
oscillatory central density dynamics can occur, as shown
in our detailed numerical simulations.

Furthermore, in a broad array of gain/loss parametric regimes, we have identified
the emergence of transient, spatiotemporally localized waveforms
reminiscent of the Peregrine rogue wave (PRW) of the integrable NLS limit.
Indeed, in a diverse array of scenarios, we have 
  exhibited temporal, as well as spatial growth and decay rates
  that are highly proximal to PRWs for finite time intervals.
  This strongly suggests that the Peregrine structures remain
  highly relevant for the dynamics beyond their ``exact'' realm
  of the integrable limit, as it was first reported in \cite{Yang1}, for the case of
  non-integrable higher-order NLS. That being said, upon sufficient departure
  from that parametric range, the gain/loss appear to overwhelm the dynamics
  and destroy the potential of manifestation of PRW-type structures.

This main finding, can be schematically depicted in two ways.
The first, is illustrated in the 
top panel of Fig.~\ref{figure8}: an initial condition $u_0(x,t_0)\in X_w$,
may evolve towards a waveform reminiscent of the PRW, achieving its maximum at
$t=t^*$.
The weighted subspace $X_w$ of the infinite dimensional phase space $X$
for the extended dynamical system associated to Eq.~(\ref{eq1}), describes the
algebraic spatial
localization of the initial condition. The second, can be portrayed in the
``phase diagram'' of the system shown in the bottom panel of Fig.~\ref{figure8}.
A potentially existing PRW-type solution [dashed-dotted (blue) curve],
is marked by $u_{\mbox{\tiny PS}}(x,t)$. It can be visualized as a ``homoclinic loop''
connecting the unstable background of density $P_0$.
The solution $u_{\mbox{\tiny PS}}(x,t)$ is an element in the weighted
subspace $X_{w}$, which is a potential domain of attraction for such solutions.
An initial condition $u_0\in X_w$, can be selected to start nearby the
orbit $u_{\mbox{\tiny PS}}(x,t)$. This, follows a path $\phi(t,u_0)$
[continuous (magenta) curve], which remains close to the orbit
$u_{\mbox{\tiny PS}}(x,t)$, for a {\it finite-time interval}.

In this picture, the ultimate fate of the path $\phi(t,u_0)$, depends on
the selected parametric regime for $\gamma$, $\delta$.
In the decay regime, the path $\phi(t,u_0)$ stays close to the orbit $u_{\mbox{\tiny
PS}}(x,t)$
for a finite time interval, but then diverges through the branch $D$ [continuous
(magenta) curve], to the trivial state
$u=0$. In the collapse regime, the path $\phi(t,u_0)$ stays close to the orbit
$u_{\mbox{\tiny PS}}(x,t)$ for shorter times, and then diverges towards collapse,
through the branch $C$ [dotted (red) curve]. Finally, in the limit set regime,
the path  $\phi(t,u_0)$ stays close to the orbit  $u_{\mbox{\tiny PS}}(x,t)$
for a finite interval, and then, it converges to the nontrivial attracting
set $\omega(\mathcal{B})$, through the branch $LS$ [dashed (green) curve].

The present study, opens a number of questions for future investigations.
An important one, concerns the development of analytical approximations to determine
such PRW-type solutions, or the derivation of analytical estimates
on the spatiotemporal growth/decay rates
observed numerically.
This is part of the broader
question regarding
the persistence
of rational solutions in models beyond the integrable limit, such
as the one of the NLS and, e.g., of the massive Thirring model~\cite{alejandro}.

Another interesting and relevant direction, is to explore the possibility
of generalizing this approach to other models, such as extended NLS equations
incorporating
higher-order effects \cite{devine,P6,Yang1,Yang2,themis}, or even to discrete setups
\cite{P4,P7}.
Appreciating the robustness of rogue waves under both
perturbations in initial data and perturbations in the models/parameters
is thus currently a broader theme under consideration, and relevant
conclusions will be reported in future work.

{\bf Acknowledgments.} We would like to thank the referees for valuable comments. The authors Z.A.A., G.F., D.J.F., N.I.K., P.G.K., I.G.S. and
K.V.,
acknowledge that this work made
possible by NPRP grant {\#} [8-764-160] from Qatar National Research Fund (a member of
Qatar Foundation).
The findings achieved herein are solely the responsibility of the authors.
D.J.F. and P.G.K. gratefully acknowledge the support of the ``Greek Diaspora Fellowship
Program''
of Stavros Niarchos Foundation.



\begin{thebibliography}{99}
	\providecommand{\natexlab}[1]{#1}
	\expandafter\ifx\csname urlstyle\endcsname\relax
	\providecommand{\doi}[1]{doi:\discretionary{}{}{}#1}\else
	\providecommand{\doi}{doi:\discretionary{}{}{}\begingroup
		\urlstyle{rm}\Url}\fi
	
	
	\bibitem[{Hasegawa \& Kodama(1996)}]{KodHas87} A.  Hasegawa and Y. Kodama, {\em Solitons in optical communications}
	(Oxford University Press, 1996).
	\bibitem[{Agrawal(2003)}]{Agra1} G. P. Agrawal, {\em Nonlinear Fiber Optics} (Academic Press, 2012).
	\bibitem[{Kivshar \& Agrawal(2003)}]{Agra2} Yu. S. Kivshar and G. P. Agrawal, {\em Optical Solitons: From Fibers to Photonic Crystals} (Academic Press, 2003).
	\bibitem{akbook} N. N. Akhmediev and A. Ankiewicz,
	{\it Solitons. Nonlinear Pulses and Beams} (Chapman and Hall, 1997).
	\bibitem{Gagnon} L. Gagnon, J. Opt. Soc. Am. B \textbf{10}, 469 (1993).
	\bibitem{H_Peregrine} D.~H.~Peregrine, %
	J. Austral. Math. Soc. B \textbf{25}, \ 16 \ (1983).
	
	\bibitem{kuz} E. A. Kuznetsov, Sov. Phys.-Dokl. {\bf 22}, 507 (1977).
	
	\bibitem{ma} Y. C. Ma, Stud. Appl. Math. {\bf 60}, 43 (1979).
	
	\bibitem{akh} N. N. Akhmediev, V. M. Eleonskii, and N. E. Kulagin,
	Theor. Math. Phys. {\bf 72}, 809 (1987).
	
	
	\bibitem{dt} K. B. Dysthe and K. Trulsen, Phys. Scr. {\bf T82}, 48 (1999).
	
	\bibitem{k2a} E. Pelinovsky and C. Kharif (eds.),
	{\it Extreme Ocean Waves} (Springer, New York, 2008).
	
	\bibitem{k2b} C. Kharif, E. Pelinovsky, and A. Slunyaev,
	{\it Rogue Waves in the Ocean} (Springer, New York, 2009).
	
	\bibitem{k2c} A. R. Osborne, {\it Nonlinear Ocean Waves and the Inverse Scattering
		Transform} (Academic Press, Amsterdam, 2010).
	
	\bibitem{k2d} M. Onorato, S. Residori, F. Baronio,
	{\it Rogue and Shock Waves in Nonlinear Dispersive Media}
	(Springer-Verlag, Heidelberg, 2016).
	
	\bibitem{hydro} A. Chabchoub, N. P. Hoffmann, and N. Akhmediev,
	Phys. Rev. Lett. {\bf 106}, 204502 (2011).
	
	\bibitem{hydro2} A. Chabchoub, N. Hoffmann, M. Onorato,
	and N. Akhmediev, Phys. Rev. X {\bf 2}, 011015 (2012).
	
	\bibitem{hydro3} A. Chabchoub and M. Fink, Phys. Rev. Lett. {\bf 112}, 124101 (2014).
	\bibitem{opt1} D. R. Solli, C. Ropers, P. Koonath, and B. Jalali,
	Nature {\bf 450}, 1054 (2007).
	
	\bibitem{opt2} B. Kibler {\it et al.}, Nature Phys. {\bf 6}, 790 (2010).
	
	\bibitem{opt3} B. Kibler {\it et al.}, Sci. Rep. {\bf 2}, 463 (2012).
	
	\bibitem{opt4} J. M. Dudley, F. Dias, M. Erkintalo, and G. Genty,
	Nat. Photon. {\bf 8}, 755
	(2014).
	
	\bibitem{opt5} B. Frisquet {\it et al.}, Sci. Rep. {\bf 6}, 20785 (2016).
	
	\bibitem{laser} C. Lecaplain, Ph. Grelu, J. M. Soto-Crespo, and N. Akhmediev,
	Phys. Rev. Lett. {\bf 108}, 233901 (2012).
	
	\bibitem{He} A. N. Ganshin, V. B. Efimov, G. V. Kolmakov, L. P. Mezhov-Deglin,
	and P. V. E. McClintock, Phys. Rev. Lett. {\bf 101}, 065303 (2008).
	
	\bibitem{plasma} H. Bailung, S. K. Sharma, and Y. Nakamura,
	Phys. Rev. Lett. {\bf 107}, 255005 (2011).
	\bibitem{devine} A. Ankiewicz, N. Devine, N. Akhmediev,
	Phys. Lett. A {\bf 373}, 3997 (2009).
	\bibitem{NR4Anki} A. Ankiewicz, Y. Wang, S. Wabnitz, and N. Akhmediev,
	Phys. Rev. E {\bf 89}, 012907 (2014).
	%
	%
	\bibitem{NR5Wang} L. H. Wang, K. Porsezian, and J. S. He, Phys. Rev. E {\bf 87}, 053202 (2013).
	%
	%
	\bibitem{NRbor6} Y. Yang, Z. Yan and B. A. Malomed, Chaos \textbf{25}, 103112 (2015).
	%
	\bibitem{NRbor5a} Y.  Wang,  L. Song,  L. LI  and B. A. Malomed, J. Opt. Soc. Am. B \textbf{32}, 2257 (2015).
	%
	%
	\bibitem{calinibook} A. Calini and C. M. Schober,
	pp.~31--51 in Ref.~\cite{k2a}.
	%
	%
	%
	%
	%
	%
	
	%
	%
	%
	%
	%
	\bibitem{KalimerisGarnier} J. Garnier and K. Kalimeris,
	J. Phys. A: Math. Theor. {\bf 45}, 035202 (2012).
	
	\bibitem{KMS} J. Cuevas-Maraver, P.G. Kevrekidis, D.J. Frantzeskakis, N.I. Karachalios, M. Haragus, G. James, \url{https://arxiv.org/abs/1701.06212}.
	
	\bibitem{onorato1} M. Onorato and  D. Proment,  Phys. Lett. A \textbf{376}, 3057
	(2012).
	
	\bibitem{ZO} E. Zakharov and L. A. Ostrovsky, Phys. D \textbf{238}, 540 (2009).
	
	\bibitem{brunetti} M. Brunetti, N. Marchiando, N. Berti, J. Kasparian, Phys. Lett. A \textbf{378}, 1025 (2014).
	
	\bibitem{EPeli} A. Slunyaev, A. Sergeeva and E. Pelinovsky, Phys. D \textbf{303}, 18 (2015).
	
	\bibitem{calini2012} A. Calini and C. M. Schober, Nonlinearity {\bf 25}, R99 (2012).
	%
	%
	\bibitem{NR11} A. Islas and C.M. Schober, Phys. D \textbf{240}, 1041 (2011).
	%
	
	
	\bibitem{Babis} C. Eigen, A. L. Gaunt, A. Suleymanzade, N. Navon, Z. Hadzibabic, and R. P. Smith, Phys. Rev. X \textbf{6}, 041058 (2016).
	
\bibitem{Yang1} G. Yang, L. Li and S. Jia, Phys. Rev. E \textbf{85}, 046608 (2012).
%
\bibitem{Yang2} G. Yang, Y. Wang,  Z. Qin,  B. A. Malomed,  D. Mihalache and L. Li, Phys.  Rev. E \textbf{90}, 062909 (2014).
%
\bibitem{Hirota} R. Hirota, J. Math. Phys. \textbf{14}, 805 (1973).

%
	\bibitem{P2} E. G. Charalampidis, J. Cuevas-Maraver, D. J. Frantzeskakis and P. G. Kevrekidis, \url{https://arxiv.org/abs/1609.01798}.
	%

	\bibitem[{Cazenave(2003)}]{Caz03} T. Cazenave, {\em Semilinear Schr{\"o}dinger equations},
	Courant Lecture Notes 10 (Amer. Math. Soc., 2003).
	%
	\bibitem[{Kato(1975)}]{Kato0} T. Kato, {\em Quasilinear equations of evolution with applications to partial differential equations}, Lecture Notes in Mathematics \textbf{448}, pp. 25--70 (Springer-Verlag, New York, 1975).
	%
	%
	%
	
	\bibitem{P6} V. Achilleos, A. R. Bishop, S. Diamantidis, D. J. Frantzeskakis, T. P. Horikis, N. I. Karachalios, and P. G. Kevrekidis
	Phys. Rev. E \textbf{94}, 012210 (2016).
%
		\bibitem[{Achilleos \emph{et~al.}(2015) }]{PartI}
		V. Achilleos, S. Diamantidis, D. J. Frantzeskakis, T. P. Horikis,
		N. I. Karachalios, and P. G. Kevrekidis, Phys. D
		\textbf{316}, 57 (2016).
		
%
	\bibitem{BM} G. Biondini and D. Mantzavinos. Phys. Rev. Lett. \textbf{116}, 043902 (2016).
	%
	%
	%
	%
	%
	%
	%
	%
	
	\bibitem{M4a} Strong collapse, Fig.~\ref{figure2}. Initial evolution:
\url{http://myria.math.aegean.gr/~karan/Fig4a.gif}.
	
	\bibitem{M4b} Strong collapse, Fig.~\ref{figure2}. Comparison with PRW:
\url{http://myria.math.aegean.gr/~karan/Fig4b.gif}.
	
	\bibitem{M5a} Weak collapse, Fig.~\ref{figure3}.  (a) Emergence of the first event:
\url{http://myria.math.aegean.gr/~karan/Fig5a.gif}.
	
	\bibitem{M5b} Weak collapse, Fig.~\ref{figure3}. Continued from (a):
\url{http://myria.math.aegean.gr/~karan/Fig5b.gif}.
	
	\bibitem{M7} Decay regime, Fig.~\ref{figure5}.
\url{http://myria.math.aegean.gr/~karan/Fig7.gif}
	
	\bibitem{M8} Limit-set regime, Fig.~\ref{figure6}. PRW initial condition:
\url{http://myria.math.aegean.gr/~karan/Fig8.gif}
	
	\bibitem{M9} Limit-set regime, Fig.~\ref{figure7}. MI dynamics:
\url{http://myria.math.aegean.gr/~karan/Fig9.gif}
	
	\bibitem{alejandro} A. Degasperis, A.B. Aceves, and S. Wabnitz,
	Phys. Lett. A {\bf 379}, 1067 (2015).
	
	\bibitem{themis} W. Cousins and T.P. Sapsis, 
	Phys. Rev. E {\bf 91}, 063204 (2015).
%
\bibitem{P4} Yannan Shen, P.G. Kevrekidis, G.P. Veldes, D.J. Frantzeskakis, D. DiMarzio, X. Lan, V. Radisic, 
 \url{https://arxiv.org/abs/1612.00031}. 
%
\bibitem{P7} G. P. Veldes, J. Cuevas, P. G. Kevrekidis, and D. J. Frantzeskakis. Phys. Rev. E \textbf{88}, 013203 (2013).
	
        
\end{thebibliography}
\end{document}